\documentclass{proposalnsf}
\usepackage{caption}
\usepackage{color}
\usepackage{graphicx}   
\usepackage{epsfig}
\usepackage{subcaption}
\usepackage{xcolor}
\usepackage{amsmath}    
\usepackage{verbatim}   
\usepackage{soul}

\def\rrr#1\\{\par
\medskip\hbox{\vbox{\parindent=2em\hsize=6.12in
\hangindent=4em\hangafter=1#1}}}

 \newcommand{\wpe}{\omega_{pe}}
 \newcommand{\wce}{\omega_{ce}}
 \newcommand{\wpi}{\omega_{pi}}
 \newcommand{\wci}{\omega_{ci}}

 \newcommand{\m}[1]{\mathbf #1} 

\begin{document}
\begin{center}
\pagenumbering{roman}
{\Huge{\bf Turbulent Dissipation Challenge}}\\
{\Large{\bf Problem Description}}\\*[3mm]
Tulasi N Parashar\footnote{CalTech NASA Jet Propulsion Laboratory, Pasadena, Currently at University of Delaware, Newark, DE},
Chadi Salem\footnote{Space Science Laboratory, University of California, Berkeley},
Robert Wicks\footnote{NASA Goddard Space Flight Center, Greenbelt, MD},
Homa Karimabadi\footnote{Department of Electrical and Computer Engineering,
University of California, San Diego},
S Peter Gary\footnote{Space Science Institute, Boulder}
Benjamin Chandran\footnote{University of New Hampshire, Durham, NH},
William H Matthaeus\footnote{University of Delaware, Newark, DE}
\end{center}

\noindent 
{\bf The Goal}\\*[3mm]
The goal of this document is to present a detailed description of
the goals, simulation setup and diagnostics for the "Turbulent
Dissipation Challenge" (http://arxiv.org/abs/1303.0204) as discussed
in the Solar Heliospheric \& INterplanetary Environment (SHINE)
2013 workshop, American Geophysical Union's Fall Meeting 2013 and
the accompanying antenna meeting in Berkeley.

\tableofcontents
\newpage
\pagenumbering{arabic}

\section{Motivation}
The near Earth space is a dynamic system where plasma coming from
the Sun, aka solar wind, and Earth's plasma sphere, aka magnetosphere,
interact. The dynamics of this system affects not only our speccraft
and astronauts but also wireless communications and in
some extreme cases, the electrical grids on Earth. It is very
important to have good predictive capabilities for this system. An
integral part of the system is the medium that connects Sun and
Earth: the solar wind.

The solar wind is observed to be turbulent. By this we mean that the
energy at the largest scales is ``cascaded" to smaller scales via
nonlinear interactions creating a broadband spectrum of incoherent
fluctuations in fields and flows.  The distribution of energy at
various scales usually follows a power law (e.g.  \cite{KolmogorovANSD41}).
The solar wind magnetic energy spectrum obeys the Kolmogorov power
law in the ``inertial range" (e.g. \cite{ColemanApJ68, MatthaeusJGR82,
GoldsteinARAA95}), which covers scales larger than the proton inertial length
$d_i = c/\wpi$ ($\sim \rho_i$) where $c$ is the speed of light and $\wpi$ is the
proton plasma frequency. At scales close to $d_i$ a break in the
spectrum occurs and it becomes steeper (e.g. \cite{SahraouiPRL09,
AlexandrovaApJ12}), indicating a loss of energy at the proton scales.
This is consistent with the observations of proton temperature being
larger than that expected from a purely adiabatic expansion (e.g.
\cite{WangJGR01}).

Exactly how the energy in flow and electromagnetic fluctuations is
converted to the thermal degrees of freedom of the protons is still
debated. The mechanism has to be largely collisionless and has to explain
features observed in the solar wind, e.g. anisotropic heating of
the protons (e.g. \cite{MarschLRSP06} and references therein).
We outlined the proposed mechanisms and the related debate in our
earlier white paper and will only summarize them here for the sake
of brevity. For a detailed discussion please refer to
\cite{ParasharArXiv13}.

The important processes proposed for explaining the anisotropic
heating of protons can be broadly divided into two categories, wave
damping mechanisms and non-wave mechanisms.

Wave damping mechanisms include cyclotron damping (e.g. \cite{HollwegJGR02}
and references therein), Landau damping (e.g. \cite{SchekochihinApJS09}
and references therein). The non-wave mechanisms involve either low
frequency intermittent structures like current sheets and reconnection
sites (e.g.  \cite{SundkvistPRL07, ParasharPP09, ParasharPP11,
OsmanApJL11, WanPRL12, KarimabadiPP13}) or ideas like stochastic
heating (e.g. \cite{ChandranApJ10-1,XiaApJ13, BourouaineApJ13}).
There are problems associated with these mechanisms that make it
difficult to propose one of the mechanisms as the dominant mechanism
for heating solar wind protons. Cyclotron heating requires sufficient
energy at high $k_\parallel$ but plasma turbulence with a mean
magnetic field (like the solar wind) becomes highly anisotropic
with most of the energy residing in high $k_\perp$. There is still
no consensus on how to provide energy to high enough $k_\parallel$
for this mechanism to be valid. Landau damping inherently produces
high $T_\parallel$ and hence would provide parallel proton heating
instead of perpendicular heating. Intermittent structures have been
shown to have a strong correlation with highly anisotropic heating
but the cause-effect relation is not completely understood and the
exact details of how the heating takes place are debated.

We anticipate that most of these processes should be active in solar
wind at any given moment. The fundamental problem of interest is
to quantify the contributions of various processes under given solar
wind conditions. The purpose of ``Turbulent Dissipation Challenge"
is to bring the community together and take a decisive step in this
direction. This is a particularly hard problem and there are many
caveats that need to be considered before making a full attempt
at addressing this issue.

\section{Definition of Goals}
The underlying issue that has hindered a constructive progress in answering the
question of dissipation in solar wind is a lack of cross comparison between
results from various studies.

\begin{itemize}
\item Most of the numerical studies are done using models that are vastly
different in their underlying assumptions and numerical schemes. E.g. electron
magnetohydrodynamics (EMHD) is the fluid desription of the electron dynamics
considering protons to be an immobile neutralizing background, hybrid particle
in cell (hybrid-PIC) treats protons as particles and electrons as a neutralizing fluid
(usually massless and isothermal) whereas gyrokinetics is a Vlasov Maxwell
system in which the gyro motion of the particles has been averaged out from the
system. These are just three examples from many possible models e.g. MHD, Hall MHD,
Hall-FLR MHD, EMHD, RMHD, ERMHD, Landau Fluid, Hybrid PIC, Hybrid Eulerian
Vlasov, Full PIC, Full Eulerian Vlasov, Gyrokinetics. 

Most of the times, the studies performed using these different models do not
have the same initial conditions or even same paramters. Hence it is not
possible to cross compare the findings of these studies.

\item Similar problems are present with spacecraft data analysis. The data can
be chosen from many spacecraft, e.g. WIND, CLUSTER, ACE, Helios, Ulysses,
Voyager to name a few. Even if the data is from the same spacecraft, the
intervals chosen could be vastly different ranging from fast wind to slow wind
to all inclusive.

Also, the analysis techniques have not been methodically benchmarked against
other analysis techniques and simulations
\end{itemize}

The above mentioned problems call for a systematic cross comparison study where
different simulation models are used to study the exact same initial conditions
under as close physical and numerical parameter regimes as possible. Artificial
spacecraft data from these simulations can then be used to benchmark different
spacecraft data analysis techniques with simulations as well as with each other.

The first step of the endeavor is to understand how well different codes capture
the essenstial underlying physics. As discussed above, the essential ingredients
of plasma turbulence that are expected to play a significant role are
intermittent structures and wave damping. Hence we define the first step of this
endeavor to be the following:

\noindent
{\bf
With given initial conditions and physical parameters, how do different
simulation models cross compare in capturing the physics of: i) intermittent
structures and ii) wave damping?
}

To do so, multiple simulation codes will perform two problems with the exact
same physical parameters and as close numerical parameters as possible. The two
problems will be:
\begin{itemize}
\item First problem will be designed to generate strong turbulence with
intermittent structures.
\item Second problem will be designed to have a spectrum of waves with a given
damping rate.
\end{itemize}

By performing these simulations using different simulation models and performing
the same set of diagnostics, we will be able to quantitatively cross compare the
results from different models. This will help us better understand how well
different codes capture the physics of interest. Artificial spacecraft data will
be provided to observers in order to establish a system that will be used to
address physics questions at a later stage in the challenge.

We now go on to describe the problem parameters and initial conditions in
detail.

\section{Problem Description}
The conditions for the first simulations will be chosen to represent solar wind
at 1 AU. Hence a plasma $\beta \sim 0.6$ will be used along with $T_e = T_i$,
$m_e/m_i=0.01$, $dt=10^{-3}\wci^{-1}$. To keep the computational costs
down, the simulations will be performed in 2.5D, i.e. the simulation
dynamics will be in a plane but with 3D components for all the
vectors. 

We understand that full 3D simulations with out of plane couplings are
required for a complete description of the dynamics, but given the computational
enormity of these models, large 3D kinetic simulations will not be
possible. A compromise between inclusion of out of plane couplings and
Reynolds number of the system (proportional to the system size) has to
be made. 

Observations show that the anisotropy and intermittency grow with the width of
the inertial range e.g. \cite{WicksMNRAS10, WicksApJ12, GrecoGRL08, WuApJL13}.
Hence, we choose to work with systems with a reasonable Reynolds
number to compare the codes. After the code comparison, the critical simulations
can be better designed in 3D.

\subsection{Intermittent Structures}
These simulations will study the intermittent structures that
emerge from the nonlinear development of a Kelvin-Helmholtz instability
(e.g. \cite{ChandrasekharBook61, MiuraJGR82, KarimabadiPP13}. The
Kelvin Helmholtz instability gives rise to large scale vortices and
current sheets. As the vortices roll up, the current sheets get
thinner and give rise to secondary tearing instabilities. This
generates a turbulent "soup" of current sheets ranging in scales
from proton to electron scales. The ease of setting up KH and the
broad range of turbulent current layers generated by it make it an
ideal candidate for studying how well different codes capture
intermittent physics.

We will follow the setup used in \cite{KarimabadiPP13} with a slight
modification for this test. The initial density $n_0$ and magnetic
field $\m{B}_0$ will be uniform. $\m{B}_0$ will be inclined w.r.t.
the plane of simulation such that $\m{B} = B_0 [\m{e}_y \sin{\theta}
+ \m{e}_z \cos(\theta)]$ with $\theta=2.86^\circ$. To allow for use
of periodic boundary conditions, we will use a double shear layer
of flow. The shears will be present at $y=0.25 L_y, 0.75L_y$ with the
shear layer flow defined as $\m{U} = U_0tanh(2\pi(y-0.25)/2L_V)\m{e}_y
$ for the layer at $y=0.25 L_y$ and $\m{U} =
-U_0tanh(2\pi(y-0.75)/2L_V)\m{e}_y$ for the layer at $y=0.75 L_y$ where
$U_0 = 10 V_A^*$, $V_A^* = B_0\sin(\theta)/\sqrt{4\pi n_0}$ and
$L_V = 4d_i$, $\beta = 0.6$, $\wpe/\wce = 1.5$. The system size
will be $L_x = 125. d_i$, $L_y = 125. d_i$, $N_x = 2048$, $N_y =
2048$. Also a perturbation of the form $\delta\m{U} = \delta U_0
\sin(0.5y/L_V)\exp(-x^2/L_V^2)$, where $\delta U_0 = 0.15 U_0$,
will be added in the shear layer to expedite the growth of the
instability.

For PIC codes, excess particles will be loaded into the transition layer to
balance the electric field because of the cross field flow $\m{E} =
(B_0U_0/c)\tanh(x/L_V)\m{e}_x$.

\subsection{Wave Physics}
If linear Vlasov theory can provide a guidance to the nature of fluctuations at
kinetic scales, we can have two main possibilities: the fluctuations could behave
predominantly like Kinetic Alfv\'en Waves or like whistler waves. It is
generally agreed upon that the fluctuations close to proton inertial scales
behave like KAWs e.g. \cite{LeamonJGR99, BalePRL05, SahraouiPRL10, HeApJL12,
SalemApJL12, TenbargeApJ12}. However whether KAWs dominate down to electron
scales or not is debated e.g. \cite{HowesPRL08, SahraouiPRL09, PodestaApJ10,
GaryGRL08, ShaikhMNRAS09}.

The Turbulent Dissipation Challenge addresses the dissipative processes at the
proton scales. Also, at the first step, the emphasis is on comparing the
capability of different simulation codes to capture wave physics. Hence, for
comparing wave physics, the initial condition will be a spectrum of Kinetic Alfv\'en
Waves (KAWs). The idea is to study how well do different codes capture wave
dynamics. The critical simulations to be done at later stages will have more
realistic composition of fluctuations.

For the first step, we propose following constraints/ requirements on the problem:
\begin{itemize}
\item As the challenge is focussed on the proton inertial scales, the initial
condition should be a spectrum of KAWs around $kd_i \sim 1$.
\item The box should have at least a decade in the inertial range above proton
scales and should also resolve at least a decade below the proton inertial
length.
\item The problem should be doable in reasonable time on modern computing
clusters.
\end{itemize}

Given the above constraints, $L_x = L_y = 125. c/\wpi$, $N_x = N_y = 2048$, will
give us $\Delta x = 0.061$ corresponding to about 16 grid points across $d_i$
and 3 grid points across $2d_e$. This will also correspond to the spectral range
of $k_{min}=0.05$ and $k_{max}=102.94$. To decide on the values of $|k|$ to
be used in the initial spectrum, we look at the dispersion curves for KAWs,
calculated using linear vlasov theory (e.g. \cite{GaryJPP86, GaryBook}), with
the parameters: $\beta=0.6$, $V_A/c = 1/15.$, $m_e/m_i = 0.01$, for a few
different angles of propagation.
\begin{figure}[!hbt]
\centering
\includegraphics[width=11cm]{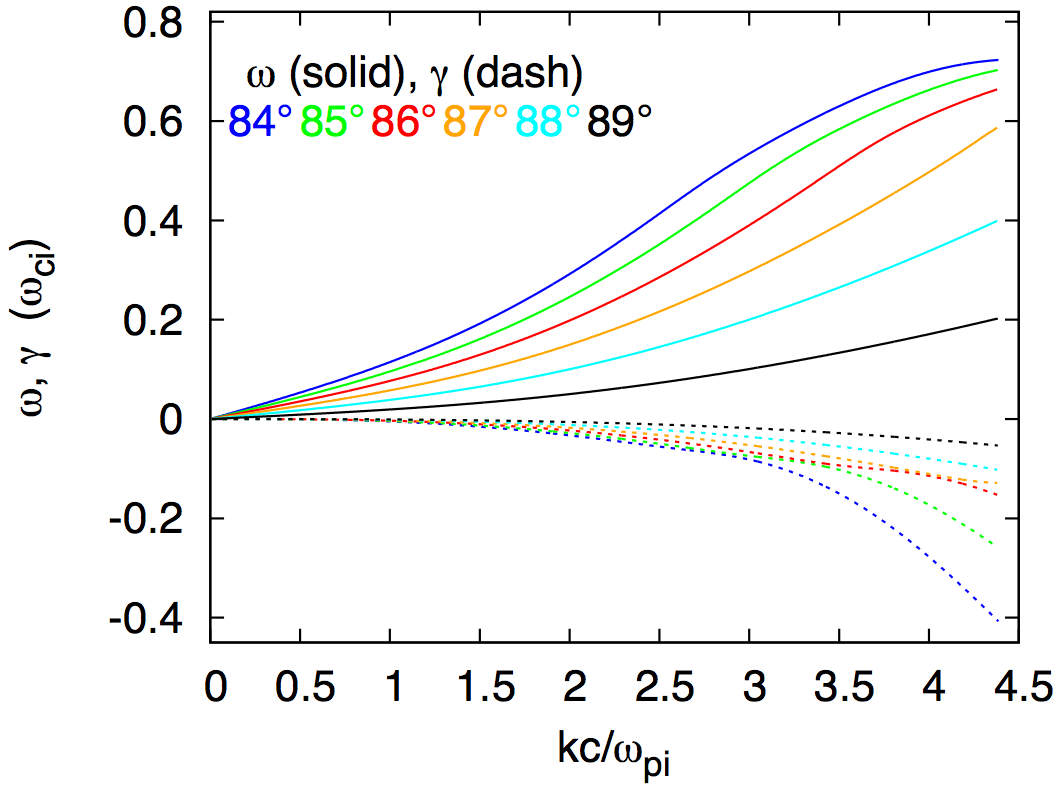}
\caption{Dispersion curves for KAWs propagating at various angles in a proton
electron Maxwellian plasma with $\beta=0.6$, $T_e = T_i$, $m_e/m_i = 0.01$,
$V_A/c = 1/15.$. Solid lines represent real freuency and dashed lines represent
the damping.}
\label{outean}
\end{figure}

Figure \ref{outean} shows the dispersion curves for various angles of
propagation of KAWs in a Maxwellian proton-electron plasma with $\beta=0.6$,
$T_e = T_i$, $m_e/m_i = 0.01$, $V_A/c = 1/15$. Almost all of these curves are
dispersive at $kc/\wpi \sim 1.$ and have very low and comparable damping rates.
We choose the window $0.5 \le k \le 1.5$ which will give us not only dispersive
waves, similar damping rates but also $\sim 66$ grid points across the smallest
wavelength in the initial spectrum. This means that the shorter wavelengths
generated by the cascade will have sufficient resolution across them.

The next constraint on the choice of $|k|$ values is that the $k_\parallel$ and
$k_\perp$ should be excitable on discrete grid points for periodicity. With this
in mind, we choose the following 8 numbers for $k_x \equiv k_\parallel$, $k_y
\equiv k_\perp$. The table shows values in units of $kd_i$ as well as number of
wavelengths in the box in parenthesis.
\begin{center}
\begin{tabular}{|cccc|}
\hline 
$k_xd_i$ & $k_yd_i$ & $|k|$ & $\theta_{kB}$\\
\hline 
0.05 (1)& 0.50 (10)& 0.0502  & 84.8294 \\
0.05 (1)& 0.65 (13)& 0.6519  & 85.6013 \\
0.05 (1)& 0.80 (16)& 0.8016  & 86.4237 \\
0.05 (1)& 0.95 (19)& 0.9513  & 86.9872 \\
0.05 (1)& 1.10 (22)& 1.1011  & 87.3974 \\
0.05 (1)& 1.25 (25)& 1.2510  & 87.7094 \\
0.05 (1)& 1.40 (28)& 1.4009  & 87.9546 \\
0.05 (1)& 1.55 (31)& 1.5508  & 88.1524 \\
\hline
\end{tabular}\\
{The parallel ($k_x$) and perp ($k_y$) wavenumbers in units of $kd_i$ as
well as the number of wavelengths in the box (in parenthesis). Corresponding
$|k|$ and $\theta_{kB}$ are also shown.}
\end{center}

The initial condition will be a spectrum of KAWs with above mentioned $|k|$ and
$\theta_{kB}$ created using the prescription in section 3 of \cite{GaryJGR04-1}.
The initial field fluctuations are written as:
\begin{equation}
\delta\m{B}(x,t=0) = \sum_{\alpha=x,y,z} \hat{\m{e}}_\alpha (\delta B_\alpha)_0
\sin(k_xx + k_yy + \phi_{B\alpha})
\end{equation}
\begin{equation}
\delta\m{E}(x,t=0) = \sum_{\alpha=x,y,z} \hat{\m{e}}_\alpha (\delta E_\alpha)_0
\sin(k_xx + k_yy + \phi_{E\alpha})
\end{equation}
\begin{equation}
\delta\m{v}_j(x,t=0) = \sum_{\alpha=x,y,z} \hat{\m{e}}_\alpha (\delta
      v_{j\alpha})_0
\sin(k_xx + k_yy + \phi_{vj\alpha})
\end{equation}

\noindent where the individual components are provided by the linear
Vlasov code (\cite{GaryJPP86}). The phases are chosen randomly.  We
expect the kinetic codes to quickly adjust the phases to the real
phases of the fluctuations.  The total amplitude of the fluctuations
will be such that $|\delta B|^2/B_0^2 \sim 0.1$. The wave vectors
output from the linear Vlasov code (\cite{GaryJPP86}) for the given
parameters are listed in the appendix.

\section{Common Diagnostics}
To facilitate a quantitative comparison between different models, a
common set of diagnostics will be performed on the simulations. Below
we outline the potential diagnostics that are reasonably
straightforward to implement. A few diagnostics common to all the simulations would be:

\begin{itemize}
\item All kinetic simulations will plot change in thermal energy of protons
as a fraction of free energy available at the beginning of the
simulation. Where possible, anisotropy as defined by
$T_\perp/T_\parallel$, with $\perp$ and $\parallel$ defined
w.r.t. mean field, should also be plotted. As an example, figure
\ref{EOS_Therm} shows the change in thermal energy as well as the
anisotropy for three hybrid simulations of Orszag-Tang vortex (OTV) with
different equation of state for electrons (for details see \cite{ParasharPP14}).

\begin{figure}
\includegraphics[width=\textwidth]{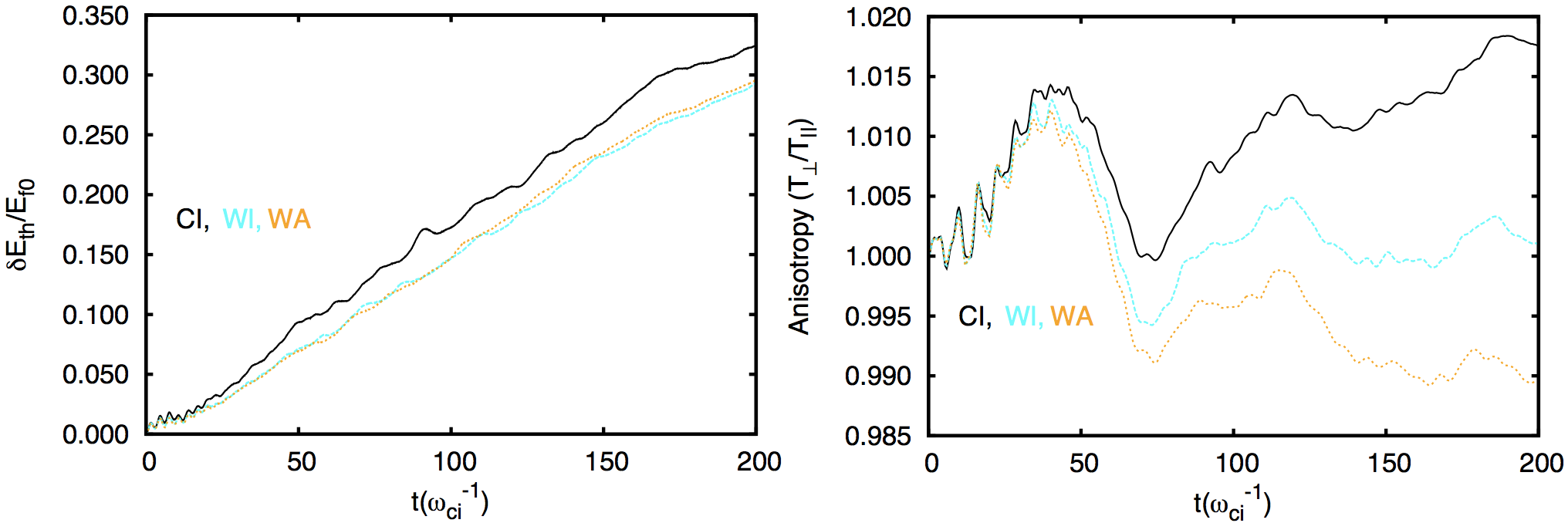}
\caption{Change in thermal energy as fraction of initially available
free energy as well as the anisotropy of proton heating for three
hybrid simulations of OTV with: cold isothermal (CI), warm isothermal (WI)
and warm adiabatic (WA) electrons. The system is turbulent after
$t\sim 70$. In the turbulent regime, $\delta E_{th}/E_{f0}$ is almost
the same for all three runs but the anisotropy is significantly
different. From \cite{ParasharPP14}, Copyright American Institute of Physics.}
\label{EOS_Therm}
\end{figure}

This way of plotting makes the results independent of the units used
in the simulation code and hence a cross comparison of proton heating
in different codes. We now describe the measures for intermittency and
wave physics.

\item Omnidirectional spectra of magnetic field in the plane perpendicular to the mean field, i.e. $E_b(k_\perp)$ where $E_b$ is total magnetic energy in shells around $k_\perp$, should also be compared from all the models. 
\end{itemize}

\subsection{Intermittent Structures}
There are multiple different measures used to quantify intermittency
of turbulence e.g. kurtosis of derivatives, scale dependent kurtosis,
filtered kurtosis, probability distribution functions (PDFs) of
increments, local intermittency measure (LIM), phase coherence index
and partial variance of increments (PVI). Out of these, we will use
scale dependent kurtosis and PDFs of increments. We describe these in
a little more detail.

\subsubsection{PDFs of Increments}
The probability distribution functions (PDFs) of turbulent quantity
are gaussian (e.g. \cite{FrischBook}) but the PDFs of increments of a
turbulent quantity are not gaussian. By taking the increments, say for
magnetic field, $\m{B}(s+\delta s) - \m{B}(s)$ we highlight the
gradients (and hence the intermittent structures) in the quantity of
interest. The strength of the gradients highlighted depends on the lag
$\delta s$. Hence the increment for a smaller lag $\delta s$
represents stronger gradients and hence most intermittent
structures. When $\delta s$ becomes comparable to the correlation
length of the system, the PDFs revert back to gaussianity. 

It has been shown that the non-gaussian tails on the PDFs of
increments correspond to the number of intermittent structures
(e.g. \cite{GrecoGRL08, SalemApJ09, GrecoApJL09, WanPP10}). Hence by comparing PDFs
of a fixed increment ($\delta s \sim 1 d_i$), from multiple models, we
can compare the number of intermittent structures resolved in each
model. This way,  we can have a quantitative measure of intermittency
as captured by different computational models.

\subsubsection{Scale Dependent Kurtosis}
The $n^{th}$ order moment of the PDFs of a variable $\phi$, also called Structure
Functions, are defined as:
\begin{equation}
S_{\phi}^n(\ell) = \left<|\phi(x+\ell) - \phi(x)|^n\right> = \left<\delta\phi_{\ell}^{n}\right>.
\label{StrFn}
\end{equation}
The Kurtosis, the $4^{th}$ order moment or $4^{th}$ structure function, measures the
Flatness of the PDFs of this given function $\phi$.  It is defined as:
\begin{equation}
\chi_{\phi} = \left<\delta\phi_\ell^4\right>/\left<\delta\phi_\ell^2\right>^2.
\end{equation}

where $\phi$ is the quantity of interest, usually the current
$\m{j}=\nabla\times \m{B}$ or the vorticity $\omega=\nabla\times\m{v}$.
It measures departures from Gaussianity. For a gaussian function
its values is equal to 3 and is more for non-gaussian functions.
Greater non-gaussianity is expected with smaller $\ell$. A higher
value of kurtosis reveals the presence of sharper concentrations
of coherent structures.  This is a straightforward diagnostic to
implement on the simulation data and being a dimensionless ratio,
should be independent of the units/ normalizations used in the
simulation codes.

\subsubsection{Structure Functions and their scaling exponents}
Within the inertial range, the structure functions defined in Eq.\ref{StrFn} scale as power laws,
\begin{equation}
S_{\phi}^n(\ell) = \left<\delta\phi_{\ell}^{n}\right> \propto \ell^{\zeta_n}
\end{equation}

The scaling exponents $\zeta_n$ are classic measures of
intermittency in hydrodynamic and MHD turbulence. 
Estimating the kurtosis (defined above) requires calculating structure
functions up to the fourth order. When it is numerically feasible,
it would be worthwhile to go up to sixth order and measure the first
six scaling exponents for v, B, and n over the scale range $2d_i <
\ell < 20d_i$ in order to provide a more complete description of
the intermittent fluctuations. Measuring these scaling exponents
will also make it possible to tie into a larger literature on
intermittency (e.g. \cite{ BrunoLRSP05, SalemApJ09,
ChandranIntermittency14}). Structure Functions are also widely used
in the analysis of solar wind fluctuations from spacecraft data,
for both inertial range (e.g. \cite{SalemApJ09}) and dissipation
range fluctuations (e.g. \cite{OsmanApJL14}).

It will be important recognize that comparisons between codes and
different runs is expected to become increasingly difficult for
higher orders n. This is due not only to intrinsic statistical
requirements, but also because higher order statistics (including
kurtosis) are expected to vary with system size (see e.g.,
\cite{AnselmetJFM84}).

\subsection{Wave Physics}
The diagnostic to be used for wave physics will compare various wave
properties. A few suggestions are:

\subsubsection{Dispersion Analysis}
\begin{figure}[!htb]
\includegraphics[width=\textwidth]{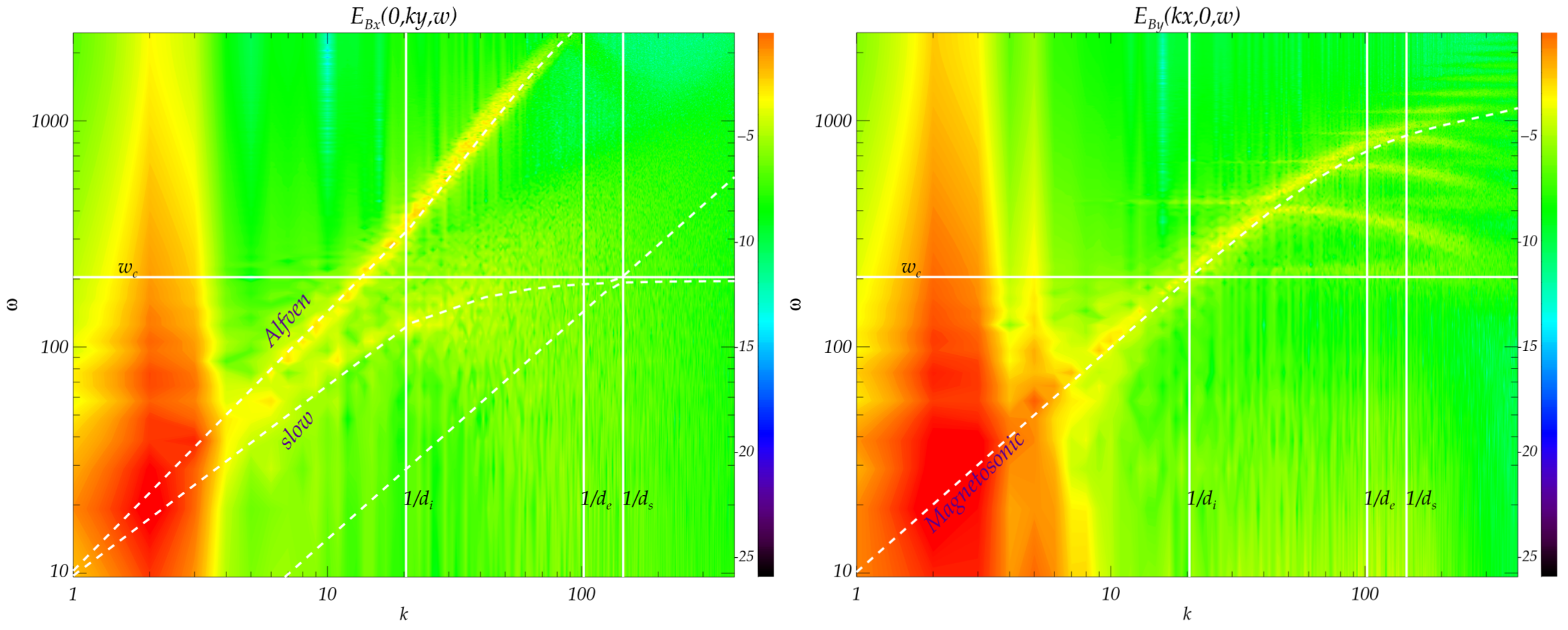}
\caption{$k,\omega$ spectra of magnetic energy from a 2.5D hybrid PIC
simulation, $\beta=0.04$ driven at large scales of the system. The
normal modes of the system are excited and we see excess energy along
the dispersion curves of Alfv\'en, slow and magnetosonic modes. From 
\cite{ParasharPhDThesis}}
\label{kw}
\end{figure}

An obvious test for the presence of waves is to look for the
appropriate dispersion in the energy spectrum. A $k-\omega$ dispersion
analysis of the simulation data will show excess of undamped energy
along the dispersion curves of the normal modes of system. As an
example figure \ref{kw} shows the spectrum of magnetic field as a
function of $k_\parallel,\omega$ and $k_\perp,\omega$ from a 2.5D
hybrid simulation with $\beta=0.04$, large system size, mean field in
the plane of simulation and driven at $|k|=2,3$ (from
\cite{ParasharPhDThesis}). The two-fluid dispersion curves for
parallel and perpendicular propagation have been
over-plottted. Outside the driving wave-numbers, there is clearly
excess energy along the dispersion curves. 

Hence $k,\omega$ diagrams are an important diagnostic to study the
normal modes that survive in a simulation's evolution.

\subsubsection{Damping Rate}
One of the most important features to cross check is if the damping
rate of the waves is appropriately captured by the simulation
codes. The spectrum of waves has been chosen such that the damping
rate is approximately the same for all the normal modes included in
the initial condition. A comparison of numerically calculated damping
rates to the average damping rate calculated from linear Vlasov theory
is an important consistency check. 

\subsubsection{Compressibility}
Two distinct compressibilities can be defined for a plasma
(e.g. \cite{GaryJGR09}). Compressibility of $j$th plasma species
defined by:  
\[C_j(\m{k}) \equiv \frac{|\delta n(k)|^2}{n_0^2}\frac{B_0^2}{|\delta
  \m{B}(k)|^2}\]

For a quasi-nerutral plasma consisting of only electrons and protons,
$C_e \sim C_p$, hence the compressibility of a only a single species
can be considered. Along with plasma compressibility, the
compressibility of magnetic field can also be computed:
\[ C_\parallel (\m{k})\equiv \frac{|\delta B_\parallel
  (\m{K})|^2}{|\delta\m{B}(\m{K})|^2}\] 

The compressibility of the dominant modes (Alfv\'en-Kinetic Alfv\'en
\& Magnetosonic-Whistler) can be calculated from linear Vlasov
theory and compared to simulations and observations (e.g.
\cite{GaryJGR09, SalemApJL12}). A direct comparison of the
compressibility values expected from linear theory with the
compressibilities captured by the simulations codes can serve as a
test of code's capcbility of capturing the wave physics.

Indeed, the compressibility is a great test parameter to use for
the identification of the dominant fluctuations or modes in the
plasma.  For instance, the compressibility of the Whistlers is
constant in this range of scales/frequencies and depends on the
angle of propagation of the mode (0 for parallel angles to 1 for
quasi-perpendicular angles).  As for the KAWs, the compressibility
increases through the dissipation range, above the ion gyro-scale
(eg. \cite{SalemApJL12}).

\subsubsection{The Electric to Magnetic Field Ratio}
The Electric to Magnetic Field ratio, $|\delta{\bf E}|/|\delta{\bf
B}|$ (\cite{SalemApJL12}), provides a complementary test for wave-mode
identification purposes. The $|\delta{\bf E}|/|\delta{\bf B}|$ ratio
is constant in the inertial range as the electric and magnetic field
fluctuations are well correlated (Bale et al. 2005). Indeed, the
electric field E in this range is essentially equivalent to ${bf
v} \times {\bf B}$ (due to the incompressible nature of the Alfvenic
fluctuations). Above the ion gyro-scale, $|\delta{\bf E}|/|\delta{\bf
B}|$ increases as a function of $k$ due to the dispersive nature
of the modes. In linear theory, $|\delta{\bf E}|/|\delta{\bf B}|$
depends on the propagation angle of the mode, both in the inertial
and dissipation range. \\

Like for the compressibility, a direct comparison of the $|\delta{\bf
E}|/|\delta{\bf B}|$ expected from linear Vlasov-Maxwell theory
with the $|\delta{\bf E}|/|\delta{\bf B}|$ captured by the simulations
will be a useful tool to understand the nature of the fluctuations.
Simulations show a very similar behavior of the $|\delta{\bf
E}|/|\delta{\bf B}|$ obtained from observations or from linear
theory.  Quantifying those similarities and differences will be
interesting.

\subsection{Comparison to Observations}
Although at this initial stage, we can not expect to see a direct comparison
with realistic observations, it will still be instructive to take artificial
cuts across simulations and compare the results to solar wind observations.
These "artificial observations" can be provided to observers for comparisons
with solar wind observations. Quantities like magnetic power spectra,
compressibilities as well as correlations between different quantities such as
reduced energy and cross helicity calculated from artificial observations can be
compared to real observations.

This will help set up the communication between simulation modelers and the
observers for future "critical simulations".

\section{Conclusion}
A quantitative improvement in our understanding of the dissipative processes active in the solar wind requires as a first step, an understanding of the best available simulation models. To do so, we propose as a first step, a comparison of the capability of these codes to capture the essential underlying physics of collisionless turbulent plasmas.

We propose to compare how well different simulation models capture intermittency and also wave damping (where applicable). A set of initial conditions and common diagnostics have been proposed that will enable a quantitative cross-model comparison.

Based on the outcome of these simulations, we can decide on doing some ``critical simulations". These simulations will be the largest possible simulations with initial conditions designed to mimic real solar wind conditions. The data obtained from these simulations will be made open to the public for doing a quantitative comparison between different dissipative processes.

\appendix
\section{Wave vectors for the KAW runs}
\begin{verbatim}

-- Dispersion
   ka_i  kc/wpi theta om_r/Om_ci g/Om_ci   |zero|  om/kvA  g/om_r
   0.275  0.502 84.83  0.0462  -5.511E-04  4.40E-12 0.0919 -0.012 
   0.357  0.652 85.60  0.0519  -1.020E-03  1.68E-10 0.0796 -0.020 
   0.439  0.802 86.42  0.0531  -1.534E-03  1.19E-10 0.0663 -0.029 
   0.521  0.951 86.99  0.0546  -2.144E-03  6.29E-11 0.0574 -0.039 
   0.603  1.101 87.40  0.0563  -2.844E-03  2.44E-11 0.0512 -0.050 
   0.685  1.251 87.71  0.0583  -3.626E-03  6.85E-12 0.0466 -0.062 
   0.767  1.401 87.95  0.0606  -4.481E-03  1.31E-12 0.0432 -0.074 
   0.849  1.551 88.15  0.0631  -5.402E-03  2.12E-09 0.0407 -0.086 

-- Electric Field
  kc/wpi  theta  (E_x/Etot)^2 (E_y/Etot)^2 (E_z/Etot)^2 (El/Etot)^2 (Etot/Btot)^2
  0.5025  84.83   2.527E-04   9.997E-01   1.791E-05   9.910E-01   4.162E-03
  0.6519  85.60   3.231E-04   9.996E-01   3.656E-05   9.930E-01   4.028E-03
  0.8016  86.42   3.411E-04   9.996E-01   5.481E-05   9.950E-01   3.876E-03
  0.9513  86.99   3.609E-04   9.996E-01   7.627E-05   9.961E-01   3.713E-03
  1.1011  87.40   3.819E-04   9.995E-01   1.003E-04   9.967E-01   3.549E-03
  1.2510  87.71   4.030E-04   9.995E-01   1.258E-04   9.971E-01   3.394E-03
  1.4009  87.95   4.225E-04   9.994E-01   1.514E-04   9.974E-01   3.258E-03
  1.5508  88.15   4.382E-04   9.994E-01   1.754E-04   9.977E-01   3.151E-03

-- E/B
  kc/wpi theta (E_x/Btot)^2 (E_y/Btot)^2 (E_z/Btot)^2 (E_k/Btot)^2 (E_kB/Btot)^2
  0.5025  84.83   1.052E-06   4.161E-03   7.455E-08   4.124E-03   3.647E-05
  0.6519  85.60   1.302E-06   4.027E-03   1.473E-07   4.000E-03   2.690E-05
  0.8016  86.42   1.322E-06   3.875E-03   2.125E-07   3.857E-03   1.822E-05
  0.9513  86.99   1.340E-06   3.712E-03   2.832E-07   3.698E-03   1.332E-05
  1.1011  87.40   1.355E-06   3.547E-03   3.559E-07   3.537E-03   1.031E-05
  1.2510  87.71   1.368E-06   3.392E-03   4.270E-07   3.384E-03   8.329E-06
  1.4009  87.95   1.377E-06   3.256E-03   4.933E-07   3.250E-03   6.977E-06
  1.5508  88.15   1.381E-06   3.149E-03   5.527E-07   3.144E-03   6.025E-06

-- B
  kc/wpi  theta (B_x/Btot)^2 (B_y/Btot)^2 (B_z/Btot)^2 (B_z/E_z)^2
  0.5025  84.83   9.720E-01   2.276E-04   2.780E-02   3.729E+05
  0.6519  85.60   9.538E-01   2.715E-04   4.589E-02   3.116E+05
  0.8016  86.42   9.323E-01   2.633E-04   6.740E-02   3.172E+05
  0.9513  86.99   9.086E-01   2.525E-04   9.114E-02   3.218E+05
  1.1011  87.40   8.838E-01   2.397E-04   1.160E-01   3.259E+05
  1.2510  87.71   8.589E-01   2.253E-04   1.408E-01   3.299E+05
  1.4009  87.95   8.352E-01   2.099E-04   1.646E-01   3.336E+05
  1.5508  88.15   8.135E-01   1.938E-04   1.863E-01   3.370E+05

-- V (V_i - first line, V_e - second line)
  kc/wpi  (dvx/dBtot)^2 (dvy/dBtot)^2 (dvz/dBtot)^2 (dv_k/dBtot)^2 (dv_kB/dBtot)^2
  0.5025   0.85   8.212E-04   7.745E-03   6.524E-04   7.914E-03
  0.5025   1.01   2.500E-04   2.097E-01   6.623E-04   2.093E-01
  
  0.6519   0.77   9.647E-04   1.178E-02   7.953E-04   1.195E-02
  0.6519   1.04   3.089E-04   3.553E-01   8.074E-04   3.548E-01
 
  0.8016   0.68   9.231E-04   1.569E-02   7.964E-04   1.582E-02
  0.8016   1.06   3.131E-04   5.411E-01   8.085E-04   5.406E-01

  0.9513   0.58   8.800E-04   1.905E-02   7.982E-04   1.913E-02
  0.9513   1.10   3.165E-04   7.673E-01   8.104E-04   7.668E-01

  1.1011   0.48   8.366E-04   2.149E-02   7.996E-04   2.152E-02
  1.1011   1.14   3.194E-04   1.034E+00   8.120E-04   1.033E+00

  1.2510   0.39   7.949E-04   2.280E-02   7.994E-04   2.279E-02
  1.2510   1.19   3.214E-04   1.339E+00   8.120E-04   1.339E+00

  1.4009   0.30   7.566E-04   2.293E-02   7.970E-04   2.289E-02
  1.4009   1.25   3.227E-04   1.681E+00   8.098E-04   1.680E+00

  1.5508   0.23   7.234E-04   2.199E-02   7.919E-04   2.192E-02
  1.5508   1.33   3.228E-04   2.054E+00   8.051E-04   2.054E+00
\end{verbatim}

\bibliographystyle{jponew}

\end{document}